\documentstyle[12pt]{article}
\def \be  {\begin{equation}}
\def \ee  {\end{equation}}
\def \ba  {\begin{eqnarray}}
\def \ea  {\end{eqnarray}}
\def \baa {\begin{eqnarray*}}
\def \eaa {\end{eqnarray*}}
\def \bb  {\begin{thebibliography}}
\def \eb  {\end{thebibliography}}
\newcommand \ci [1] {\cite{#1}}

\def \lab #1 {\label{#1}}
\newcommand\re[1]{(\ref{#1})}
\def \qqquad {\qquad\quad}
\def \qqqquad {\qquad\qquad}
\newcommand\lr[1]{{\left({#1}\right)}}

\def \vev #1 {\langle{#1}\rangle}
\def \VEV #1 {\left\langle{#1}\right\rangle}
\newcommand \ket [1] {|{#1}\rangle}
\newcommand \bra [1] {\langle {#1}|}
\def \e {\mbox{e}}
\def \CO {{\cal O}}

\def \CM {{\cal M}}
\def \D {{\rm D}}
\def \B {{\rm B}}
\def \b {{\rm b}}

\def \c {{\rm c}}

\def \alpi {\frac \as \pi}
\def \fracs #1#2 {{\mbox{\small $\frac{#1}{#2}$}}}
\def \tt {t}
\newcommand{\ch}{\mathop{\rm ch}\nolimits}
\newcommand{\sh}{\mathop{\rm sh}\nolimits}
\newcommand{\as}{\ifmmode\alpha_{\rm s}\else{$\alpha_{\rm s}$}\fi}
\newcommand{\MS}{\ifmmode\overline{\rm MS}\else{$\overline{\rm MS}$}\fi}

\begin{document}
\def\thefootnote{\fnsymbol{footnote}}
\thispagestyle{empty}
\hfill\parbox{35mm}{{\sc OUT--4102--53}\par
                    {\sc ITP--SB--94--47}\par
                         hep-ph/9411323  \par
                         November, 1994}
\vspace*{30mm}
\begin{center}
{\LARGE Renormalized sum rules for structure functions \\[3mm]
        of heavy mesons decays}
\par\vspace*{20mm}\par
{\large A.~G.~Grozin}$\ {}^{1,}$%
\footnote{A.Grozin@open.ac.uk; on leave from Budker Institute of Nuclear
Physics, Novosibirsk 630090, Russia}
and {\large G.~P.~Korchemsky}$\ {}^{2,}$%
\footnote{Korchems@max.physics.sunysb.edu; on leave from the Laboratory
of Theoretical Physics, JINR, Dubna, Russia}
\par\bigskip\par\medskip
${}^1${\em Physics Department, Open University, Milton Keynes, MK7 6AA, UK}
\par\medskip
${}^2${\em Institute for Theoretical Physics,
State University of New York at Stony Brook, \par
Stony Brook, New York 11794 -- 3840, U.S.A.}
\end{center}
\vspace*{20mm}

\begin{abstract}
We consider properties of the structure functions of inclusive heavy meson
decays $\B\to X_c$ and treat the $\c$ quark mass as a free parameter. We
show that in two extreme cases of heavy and light $\c$ quark the structure
functions of heavy--heavy and heavy--light transitions
are given by a Fourier transform of the matrix elements of Wilson lines
containing a time--like and a light--like segment, correspondingly.
Using the renormalization properties of Wilson lines we find the
dependence of the structure functions on the factorization scale,
the structure function
of heavy--heavy transition is renormalized multiplicatively while
that of heavy-light transition obeys the GLAP--type evolution equation.
We propose a generalization of the sum rules for the moments of the
structure functions (Bjorken, Voloshin, and the ``third'' sum rules)
with a soft exponential factorization cut--off, which correctly incorporates
both perturbative and nonperturbative effects. We analyze nonperturbative
corrections by first considering infrared renormalon contributions to the
Wilson lines. Uncertainties induced by the leading renormalon pole at
$u=\frac12$ are exactly cancelled by the similar uncertainty in the
heavy quark pole mass. The leading nonperturbative corrections associated
with the next renormalon at $u=1$ are parameterized by the matrix
element $\mu_\pi^2$ which is proportional to the heavy quark kinetic energy.
\end{abstract}
\newpage

\def\thefootnote{\arabic{footnote}}
\setcounter{footnote} 0
\section{Introduction}
\lab{Intro}

Recently, the interest to the inclusive semileptonic decays of the $\B$ meson
was renewed after it was proposed~\ci{VS,CGG} to apply the operator product
expansion (OPE) in powers of a heavy quark mass to the analysis of the heavy
meson decays. Being combined together, the OPE and heavy quark effective
theory allow us to expand the total decay rates of the $B$ meson in powers of
$\Lambda/m_b$, with $\Lambda$ the QCD scale and $m_b$ the $\b$ quark mass,
and parameterize nonperturbative corrections by hadronic matrix elements%
{}~\ci{BUV}. The analysis of the differential distributions of the $\B$ meson
in the framework of the heavy quark expansion turned out to be more
complicated~\ci{dif}.
Let us take as an example the inclusive decay $\B\to W + X_c$ with
$W\to l\bar\nu$ and consider the decay distribution
${d\Gamma_{\B\to W + X_c}}/{d W^2 d W_0}$ with respect to the invariant mass
$W^2$ and the energy $W_0$ of the lepton pair in the rest frame of the
$\B$ meson. The process has two large momentum scales, the heavy quark mass,
$m_b$, and the invariant mass, $m_X$, of the final state
$$
m_X^2=(m_B v - W)^2 =m_B^2 - 2 m_B W_0 + W^2
$$
with $v_\mu$ the velocity of the $\B$ meson. The OPE of the differential
distribution~\ci{dif} is an expansion in powers of the $\b$ quark mass,
$\Lambda/m_b$, and virtuality
of the ~outgoing $\c$ quark, $\Lambda m_b/(m_X^2-m_c^2)$, provided that both
parameters are small. Although this expansion is well defined for large
invariant masses $m_X$, it becomes divergent in the
region where
\be
m_X^2-m_c^2  < \Lambda m_b\,.
\lab{kin}
\ee
In this region the OPE fails and the differential distributions in the
decay $\B\to W+X_c$ are parameterized in the leading $\Lambda/m_b$ order by
the universal nonperturbative structure function of the $\B$
meson~\ci{str}. This function takes into account large perturbative and
nonperturbative corrections to the differential rate of the decay
in the region \re{kin}. The structure function cannot be
calculated at present out of QCD, but it turns out to be possible to
{}~relate its integral characteristics to a few phenomenological
parameters. The corresponding relations are known as Bjorken%
{}~\ci{Bj,IW2}, Voloshin~\ci{Vol,Bur}, and the ``third''~\ci{BGSUV}
sum rules. They express first three moments of the structure function
in terms of the fundamental parameters of the HQET.
It was realized, however,~\ci{IW2,Vol,Bur},
that these parameters become functions of
the normalization scale after we take into account perturbative
corrections and the sum rules should be properly modified to
incorporate perturbative effects. It remains unclear how to renormalize
consistently the sum rules for the structure function and
this is one of the questions we consider in the present paper.
The properties of the structure function crucially depend \ci{str}
on the ratio of the quark masses $m_c/m_b$. We study below
the structure function of the $\B$ meson and treat the $\c$ quark mass
as a free parameter.

It is convenient to define the angle, $\vartheta$, in Minkowski space--time
between 4--velocities $v=p_B/m_B$ and $v'=p_X/m_X$ of the $\B$ meson and $X_c$,
respectively, as $\ch\vartheta = (v\cdot v')$, and perform analysis
using the variables $\vartheta$ and $m_X$,
$$
W^2=m_X^2+m_B^2-2 m_X m_B \ch\vartheta\,,\qqquad
W_0=m_B-m_X\ch\vartheta\,.
$$
Then, the phase space for $m_X^2$ and $\vartheta$ is
$$
m_D\le m_X \le m_B\,,\qqqquad
0\le \vartheta \le \ln\frac{m_B}{m_X}\,,
$$
where the maximum value of $\vartheta$ is ~achieved in the exclusive
$\D$ meson production, $m_X=m_D$. In this paper we study the $\B$ meson decay
in the subregion of the phase space defined in~\re{kin}.

Considering all possible values of $m_c$ in~\re{kin}, we distinguish two
special cases of ``heavy'' and ``light'' $\c$ quark mass,
$m_c^2 \gg \Lambda m_b$ and $m_c^2 \ll \Lambda m_b$, respectively.
The real masses $m_b$, $m_c$ lie somewhere between these two
extreme cases. In the first case, $m_c^2 \gg \Lambda m_b$,
the relation~\re{kin}
implies that the $\c$ quark is nearly on mass shell in the final state $X_c$.
Light components of $X_c$ have an energy of order $\Lambda$ and
interacting with them the $\c$ quark
behaves as a heavy quark. In particular, for $m_c^2 \gg \Lambda m_b$ the
momentum $p_X=m_X v'$ of the final state $X_c$ can be decomposed
similar to the momentum $p_B=m_Bv$ of the $\B$ meson as
\be
p_X=m_c v' + k'\,,\qqqquad
p_B=m_b v  + k\,,
\lab{dec}
\ee
with $k'$ and $k$ being the residual momenta. This allows us to apply the heavy
quark expansion in both $m_b$ and $m_c$ and introduce the notion of the
residual energy $\varepsilon_X = v'\cdot k'$ for the $X_c$ state
similar to that for the $\B$ meson,
\be
\varepsilon_X   % = \frac{m_X^2-m_c^2}{2m_c}
=m_X-m_c\,,      % + \CO(1/m_c)
\qqqquad
\varepsilon_0=m_B-m_b\,.
\lab{res}
\ee
In the limit of the ``light'' $\c$ quark, $m_c^2 \ll \Lambda m_b$,
there are two different subregions in~\re{kin}.
At $m_X^2-m_c^2=\CO(m_c\Lambda)$,
or equivalently at $\varepsilon_X\ll m_c$,
the $\c$ quark may be considered
as a heavy; we may apply HQET and obtain results for an infinitely heavy $\c$
quark plus small $1/m_c$ corrections.
In the opposite case, $m_b\Lambda \gg m_X^2\gg m_c^2$,
we have $\varepsilon_X\gg m_c$, which means that the light
components of $X_c$ have the energy much bigger $m_c$ and interacting with
them the $\c$ quark behaves effectively as massless.

\section{Heavy to heavy transitions}
\lab{HH}

In the region $\varepsilon_X\ll m_c$, the $\c$ quark can be treated as
heavy and we analyze heavy to heavy meson transition $\B\to W X_c$
in the leading heavy mass limit neglecting both $1/m_b$ and $1/m_c$
corrections. The momenta of the mesons can be decomposed as in~\re{dec}
and \re{res}
with the residual energy $\varepsilon_0$ and $\varepsilon_X$ of order
$\Lambda$.
We may switch off the spins of the $\b$ and $\c$ quarks in this
approximation, and consider the decay $b_v \to c_{v'} W$
with $b_v$, $c_{v'}$ and $W$ spinless.
Spin Clebsh--Gordan factors can be trivially included later.

\subsection{Inclusive structure function}

To simplify consideration we choose the effective interaction lagrangian
to be $gJW^\dagger+{\rm h.c.}$ with $g$ the coupling constant
and $J=c_{v'}^\dagger b_v$. Then the decay matrix elements are
$\CM(\b\to\c W)=g$ and $\CM(\B\to X_c W)=g\bra{X_c} J \ket{\B}$.
The inclusive differential rate of the decay has the form~\ci{Bj}
\begin{equation}
d\Gamma(\B\to X_c W) = d\Gamma_0(\b\to\c W)
F(\vartheta,\varepsilon) d\varepsilon
\lab{SF}
\end{equation}
with $\varepsilon$ being the energy of the light components of the state
$X_c$ in the rest frame of the $\c$ quark, and $\vartheta$
the angle between quark velocities $v$ and $v'$ in~\re{dec}.
The structure function is defined in HQET as
\be
F(\vartheta,\varepsilon) =
\sum_{X_c} \left| \bra{X_c} J \ket{B} \right|^2
\ \delta(\varepsilon-\varepsilon_{X}),
\lab{str}
\ee
where the averaging over $\B$ polarizations is assumed and the summation
is performed over the states $X_c$ in HQET with the residual energy
$\varepsilon_{X}$ defined in~\re{res}.
The states $X_c=\c\bar q$ can be classified according to the angular
momentum and parity, $j^P$, of the light component $\bar q$. Namely,
the ground state (S--wave) has $j^P=\frac12^-$, and the excited
P--wave states have $j^P=\frac12^+$ and $\frac32^+$.
Exclusive $\B\to X_c$ decays to S-- and P--wave mesons $X_c$
are described by the form factors~\ci{IW,IW2}
\baa
\bra{\fracs12 ^-,n}J\ket{\fracs12 ^-}
      &=& \xi^{(n)}(\ch\vartheta)\ \overline{u}'u\,,\quad
\\[2.5mm]
\bra{\fracs12 ^+,n} J \ket{\fracs12 ^-}
      &=& 2\;\tau_{1/2}^{(n)}(\ch\vartheta)\ \overline{u}' \gamma_5 u\,,\quad
\\[2.5mm]
\bra{\fracs32 ^+,n}J\ket{\fracs12 ^-}
      &=& \sqrt3\; \tau_{3/2}^{(n)}(\ch\vartheta)\  v_\mu\overline{u}'_\mu u\,,
\eaa
where $\ket{\fracs12 ^-}$ is the ground state $B=\b\bar q$ meson and
where $u$ and $u_\mu$ are spin $\frac12$ and $\frac32$ meson wave functions.
Substituting these relations into~\re{str} we get the ``phenomenological''
expression for the inclusive structure function in terms of the
exclusive form factors~\ci{IW2}
\ba
&&F(\vartheta,\varepsilon) =
{\textstyle\frac12}(\ch\vartheta+1) \sum_n |\xi^{(n)}(\ch\vartheta)|^2
\delta(\varepsilon-\varepsilon_{1/2^-}^{(n)})
\nonumber\\
&&\quad+2(\ch\vartheta-1) \sum_n |\tau_{1/2}^{(n)}(\ch\vartheta)|^2
\delta(\varepsilon-\varepsilon_{1/2^+}^{(n)})
\lab{SFFF}
\\
&&\quad+(\ch\vartheta+1)^2(\ch\vartheta-1) \sum_n
|\tau_{3/2}^{(n)}(\ch\vartheta)|^2
\delta(\varepsilon-\varepsilon_{3/2^+}^{(n)})+\cdots
\nonumber
\ea
where the dots denote the contribution of D-- and higher wave states.
We follow~\ci{Gr} in this simple derivation using spinless quarks and $W$.
The structure function $F(\vartheta,\varepsilon)$
vanishes identically at $\varepsilon<\varepsilon_0$,
where $\varepsilon_0=\varepsilon_{1/2^-}^{(0)}$ is the residual energy of
the ground state meson. At this point it has a $\delta$--peak due to the
ground state contribution. Then other narrow peaks follow, separated by
gaps of order $\Lambda$ and corresponding to the excited states in \re{SFFF}.
As $\varepsilon$ grows, the peaks become wider, and at
$\varepsilon\gg\Lambda$ the structure function $F(\vartheta,\varepsilon)$
becomes a smooth function of $\varepsilon$ determined by a hard gluon
emission.

\subsection{Relation to Wilson lines}

The structure function~\re{str} can be represented as a Fourier
transformed matrix element of the product of two currents,
\be
F(\vartheta,\varepsilon)=\int d^4 x \ \e^{-iW\cdot x}\
\bra{B}J^\dagger(x)J(0)\ket{B}\,,\qqquad
W=p_B-p_X\,.
\lab{fou}
\ee
The matrix element of the current $J=c_{v'}^\dagger b_v$ in~\re{fou} can
be simplified in the leading heavy quark limit. We recall that for
$\varepsilon \ll m_c$, the $\b$ and $\c$ quarks behave as heavy particles
which move along their velocities $v$ and $v'$, respectively, and interact
with light components of the meson. Moreover, in the leading power of
$1/m_c$ and $1/m_b$ the only effect of this interaction is the appearance
of the eikonal phase in the heavy quark wave functions~\ci{phase}.
This phase is given
in QCD by a Wilson line, $P\exp(i\int_C dx\cdot A(x))$, evaluated along the
classical quark trajectory $C$ with the gauge field $A_\mu(x)$ describing
soft gluons~\ci{IK}. Then, evaluating the matrix element
%$\bra{X_c}c_{v'}^\dagger(x)b_v(x)\ket{B}$,
$\bra{\B} J^\dagger(x) J(0)\ket{\B}$,
we replace the quark field operators as follows
$$
b_v(0)=\Phi_v[0,-\infty] a_v\,,\qquad
b_v^\dagger(x)=\e^{i m_b v\cdot x} a_v^\dagger \Phi_{-v}[-\infty,x] \,,\qquad
c_{v'}(x)c_{v'}^\dagger(0)=\e^{-i m_c v'\cdot x} \Phi_{v'}[x,0] D_{v'}(x)\,,
$$
where $m_b$ and $m_c$ are bare quark masses, $a_v$ and $a_{v'}^\dagger$
are heavy quark creation and annihilation operators, and
$D_{v'}(x)$ is a free cut propagator of a heavy quark,
$$
D_{v'}(x)=\int \frac{d^4 k}{(2\pi)^4}\ \delta(k\cdot v') \ \e^{i k\cdot x}
=\int_{-\infty}^\infty \frac{dt}{2\pi}\ \delta^{(4)}(x-v't)\,.
$$
The eikonal phases
$\Phi_v[x,-\infty]$ and $\Phi_{v'}[x,0]$ correspond to the motion the
$\b$ quark with velocity $v$ from $-\infty$ to the point $x$ and to the
propagation of the $\c$ quark from the point $0$ to $x=v't$ with
velocity $v'$, respectively, and are given by
$$
\Phi_v[x,-\infty]=P\exp\lr{i\int_{-\infty}^0 ds v\cdot A(x+vs)}\,,\qquad
\Phi_{v'}[v't,0]=P\exp\lr{i\int_0^t ds v'\cdot A(v's)}\,.
$$
As a consequence, we get the following expression for the matrix element
in the leading heavy quark limit,
\be
\bra{\B}J^\dagger(x)J(0)\ket{\B}=
\int_{-\infty}^\infty \frac{dt}{2\pi} \delta^{(4)}(x-v't)
\e^{-it(m_c-m_b\ch\vartheta)}\
\bra{\B}
a_v^\dagger\; \Phi_{-v}[-\infty,t v']
            \Phi_{v'}[t v',0]
            \Phi_v[0,-\infty]
a_v
\ket{\B}
\lab{eik}
\ee
Here, three Wilson lines may be represented as a single
Wilson line integrated along the path $C$ %shown in fig.1. The path
which goes
along $v$ from $-\infty$ to point $0$, then from point $0$ along $v'$ to
point $t v'$ and finally from point $t v'$ to $-\infty$ along $-v$.
Then, for the structure function we get the expression as a Fourier
transformed Wilson line expectation value
\be
F(\vartheta,\varepsilon) =
\int_{-\infty}^\infty \frac{dt}{2\pi}\, \e^{it (\varepsilon-
\varepsilon_0 \ch\vartheta)}
W_C(\vartheta,t)
\lab{fin}
\ee
with the Wilson line defined as
\be
W_C\equiv
\bra{\B(\varepsilon_0)} P\exp\lr{i\int_C dx\cdot A(x)} \ket{\B(\varepsilon_0)}
\lab{W-def}
\ee
Here, in the leading heavy quark limit the state
$\ket{\B(\varepsilon_0)}=a_v\ket{\B}$ describes only the light components of
the $\B$ meson with amputated $\b$ quark. We notice that since the structure
function has a meaning of a cross section and not of a Green function,
the gauge fields are ordered in~\re{fin} along the path~$C$ and not in time.
However, on different parts of the path~$C$, the path--ordering
automatically implies the time--ordering. According to the definition \re{fin},
the structure function $F(\vartheta,\varepsilon)$ is a universal
distribution which depends only on the properties of the $\B$ meson and
not on the particular form of the partonic subprocess and the effective
interaction lagrangian. The $\vartheta$--dependence of the
structure function comes, apart from shift of $\varepsilon$, from the
properties of the Wilson line as a functional of the integration path~$C$.

Relation~\re{fin} implies that in order to understand the properties
of the structure function out of QCD one need to know nonperturbative
estimate for the Wilson line expectation value. Let us invert~\re{fin}
and perform the Wick rotation $t\to -i\tt$,
\be
W(\vartheta,-i\tt)=\e^{\tt\varepsilon_0(\ch\vartheta-1)}
\int_{\varepsilon_0}^\infty d\varepsilon\
\e^{-\tt(\varepsilon-\varepsilon_0)}
F(\vartheta,\varepsilon)\,,
\lab{W-F}
\ee
where we took into account that $F(\vartheta,\varepsilon)$ vanishes
for $\varepsilon < \varepsilon_0$.
Then, using eq.~\re{SFFF} we can get the phenomenological expression for the
Wilson line in terms of the exclusive form factors.
Once we know the Wilson line expectation
value we could extract the structure function and the exclusive
form factors from~\re{fin} or~\re{W-F}.
For instance,
the ground state form factor is related to the asymptotic behavior of the
Wilson line at large~$\tt$. Although at present we are not able to
evaluate a Wilson line nonperturbatively, there is
limiting case, %the small velocity limit $\vartheta\to 0$ and
the small $\tt$ limit, in which we may effectively use the relation%
{}~\re{W-F} to extract an information about the transition form factors.

\subsection{Renormalization of the structure function}

Both the exclusive form factors and the inclusive structure function
depend on the normalization point $\mu$. The $\mu$ dependence is introduced
by the factorization procedure which one has to perform on the differential
rate~\re{SF} in order to separate contributions of gluons with energy bigger
and smaller than $\mu$ to the perturbative coefficient function,
$d\Gamma_0$,
and to the nonperturbative structure function, $F$, respectively.
Each of these functions depends on $\mu$, but this dependence is compensated
in the physical distribution~\re{SF}.

Relation~\re{fin} implies that the
$\mu$--dependence of the structure function follows from the renormalization
properties of the Wilson line. The same is true for the heavy quark
exclusive form factors, which are defined by the matrix elements of the
form $\bra{X_c}J(0)\ket{\B}$. In the leading heavy quark limit, these
matrix elements can be represented as expectation values of Wilson
line with integration path which goes from $-\infty$ to point $0$
along the velocity $v$ of the $\b$ quark and then from $0$ to $\infty$
along the velocity $v'$ of the $\c$ quark. The integration path has
a cusp at point $0$ with the angle $\vartheta$ and, as a consequence,
the exclusive form factors obey the RG equation~\ci{IKR}
\be
\lr{\mu\frac{\partial}{\partial\mu}
+\beta(g)\frac{\partial}{\partial g} + \Gamma_{\rm cusp}(\as,\vartheta)}
\xi^{(n)}(\ch\vartheta)=0\,,
\lab{RG1}
\ee
with the same equation for $\tau^{(n)}(\ch\vartheta)$.
Here, $\Gamma_{\rm cusp}(\as,\vartheta)$ is the cusp anomalous dimension
known up to two--loop order~\ci{KR}. It is important
for us that $\Gamma_{\rm cusp}(\as,\vartheta)$ depends on the angle
$\vartheta$ and in the small velocity limit it has the following
asymptotics~\ci{KR}:
\be
\Gamma_{\rm cusp}(\as,\vartheta)\stackrel{\vartheta\to 0}{=}
\vartheta^2 \gamma_{\rm cusp}(\as)
+ {\cal O}(\vartheta^4)\,,
\quad
\gamma_{\rm cusp}(\as)=
\frac{\as}{3\pi}C_F+\left(\frac{\as}{\pi}\right)^2
C_F\left(C_A\left(\frac{47}{54}-\frac{\pi^2}{18}\right)-
N_f\frac5{54}\right)
\lab{small}
\ee
Since the integration path $C$
has two cusps at points $0$
and $tv'$, the Wilson line $W_C$ entering into the definition of
the structure function % is renormalized multiplicatively and
obeys the renormalization group equation
\footnote{It is important to notice that the Wilson line \re{W-def}
          entering this equation does not involve the time-ordering
          of gauge fields. Had we apply the time-ordering the
          anomalous dimension should be replaced in \re{RG} by
          $\Gamma_{\rm cusp}(\as,\vartheta)+\Gamma_{\rm
cusp}(\as,i\pi-\vartheta)$.
          }
\be
\lr{\mu\frac{\partial}{\partial\mu}
+\beta(g)\frac{\partial}{\partial g} + 2 \Gamma_{\rm cusp}(\as,\vartheta)}
W(\vartheta,-i\tt) =0 \,.
\lab{RG}
\ee
We conclude from~\re{fin} that the structure function
$F(\vartheta,\varepsilon)$ obeys the same equation.
Then, the $\mu$--dependence of the structure function
and the form factors, eqs.~\re{RG} and~\re{RG1}, is
perfectly consistent with the expansion in exclusive channels~\re{SFFF}.

As follows from~\re{W-F}, $\tt$ is conjugated to the energy, $\varepsilon$,
of light particles emitted to the final state $X_c$. Large values of
$\tt$ correspond to small residual energy of the final state in
the decay $\B\to W X_c$ and, at the same time, to the ``large'' size
Wilson line \re{W-def}. This implies that the large $\tt$ limit is essentially
nonperturbative. On the other hand, for small values of $\tt$ the residual
momentum of $X_c$ is formed by a hard gluon radiation. This suggests that
the small $\tt$ behavior of the Wilson line entering~\re{fin}
and~\re{W-F} can be analyzed perturbatively.

Perturbative corrections to the Wilson line in the small $\tt$
limit do not depend on the states involved in its matrix element~\re{W-def},
so we may consider its vacuum average for simplicity.
Expanding the path--ordered exponential~\re{W-def} in powers
of the gauge field and applying the Feynman rules, we get
\begin{equation}
W_{\rm pert}(\vartheta,-i\tt)=
1+2\as C_F\mu^{4-D}\int\frac{d^Dk}{(2\pi)^{D-2}}
\delta(k^2)\theta(k_0)
\left(\frac{v}{v\cdot k}-\frac{v'}{v'\cdot k}\right)^2
\left(1-e^{-\tt(v'\cdot k)}\right)
+{\cal O}(\as^2)\,,
\lab{WC1}
\end{equation}
where the dimensional regularization with $D=4-2\epsilon$ was used.
Integration over gluon momentum gives rise to a single pole in
$1/\epsilon$. Performing renormalization of $W_{\rm pert}$ in the \MS{}
scheme, we obtain the one--loop expression for the Wilson line%
{}~\ci{KR1}
\begin{equation}
W_{\rm pert}=1-\frac{\as}{\pi} C_F\left[
2(\vartheta\coth\vartheta-1)\log\frac{\mu\tt}{2}
+2\coth\vartheta\int_0^\vartheta dx x
\coth x -1 -\vartheta\coth\vartheta \right]+{\cal O}(\as^2)\,.
\lab{one}
\end{equation}
One easily verifies that this expression satisfies the RG equation~\re{RG}.
To higher orders of perturbation theory,
the Wilson line $W_{\rm pert}$ is a dimensionless
function of $\as(\mu)$, $\mu\tt$ and the cusp angle $\vartheta$.
Then, solving the RG equation \re{RG} we get
\begin{equation}
W_{\rm pert}(\vartheta,-i\tt)=\exp\left(
-2\int_{2/\tt}^\mu \frac{dk_t}{k_t}\Gamma_{\rm cusp}(\as(k_t),\vartheta)
+\gamma(\as(2/\tt),\vartheta)\right)
\lab{pert}
\end{equation}
where $\gamma(\as,\vartheta)$ appears as an integration constant.
The expression~\re{pert} resumes all leading and nonleading logarithmic
corrections $(\as^k\log^n \tt)$ to the Wilson line $W$.
The one--loop expression
for $\gamma(\as,\vartheta)$ can be obtained from comparison of~\re{pert}
with~\re{one}:
\begin{equation}
\gamma(\as,\vartheta)=\frac{\as}{\pi} C_F\left(
1+\vartheta\coth\vartheta-2\coth\vartheta\int_0^\vartheta dx x
\coth x\right)\,.
\lab{gamma}
\end{equation}
Although
perturbation theory perfectly describes all logarithmic corrections to
$W$, but it fails to describe uniquely power corrections in $\tt$
due to the presence of infrared renormalons~\ci{IRR} in perturbative series
for the Wilson line~\ci{KS}.

\subsection{IR renormalons in Wilson lines}

The standard analysis of the IR renormalons in physical quantities
is performed in the large $N_f$ limit with $\as N_f$ fixed. In this
limit only quark loop corrections to the gluon self--energy survive,
and one assumes that their contribution qualitatively describes
the asymptotic behavior of the whole series of PT. Let us show that
this last assumption is not valid for the structure function.
The structure function gets large perturbative corrections
which are resummed into the Wilson line expectation value~\re{pert}.
Perturbative expression~\re{pert} implies that the Wilson
line exponentially decreases as the size of the integration
path, $\tt$, increases. By examining the contribution of quark loops to this
asymptotic behavior we find
that the quark loops contribute to the anomalous
dimensions $\Gamma_{\rm cusp}$ and $\gamma$ entering into~\re{pert}
starting at $\as^2N_f$ order.
As a consequence, if we represent~\re{pert} as $\exp(-X)$ then
only first two terms, $1-X$, in the expansion of the exponent survive
in the large $N_f$ limit and their asymptotics has nothing to do with
the asymptotics
of the Wilson line~\re{pert}. Thus the large $N_f$ limit is not
applicable for the analysis of the asymptotic behavior of the Wilson
line.

Instead of taking the large $N_f$ limit, we perform an ``improved''
calculation of the one--loop Wilson line~\ci{KS}.
First, the argument of the coupling constant in \re{WC1} is fixed
after one takes into account higher order corrections to $W_C$,
to be the transverse momentum of gluons, $\as=\as(k_t)$, and not the
virtuality~\ci{ABCMV}. Second, we use nonabelian exponentiation theorem%
{}~\ci{NAexp} to
rewrite the Wilson line as an exponential of the lowest--order result,
\ba
W(-i\tt,\vartheta)&=&\exp\lr{
2C_F\mu^{4-D}\int\frac{d^Dk}{(2\pi)^{D-2}}
\delta(k^2)\theta(k_0)\as(k_t)
\left(\frac{v}{v\cdot k}-\frac{v'}{v'\cdot k}\right)^2
\left(1-e^{-\tt(v'\cdot k)}\right)
}
\nonumber
\\
&\times& \exp\lr{\tt \Delta m (1-\ch\vartheta)}\,.
\lab{W-IR}
\ea
Here, the first exponent describes the interaction of the eikonal current,
created by heavy $\c$ and $\b$ quarks,
$J_{\rm eik}(k)=\frac{v}{(vk)}-\frac{v'}{(v'k)}$,
with soft gluon radiation, while the
second exponent is due to the heavy quark self--energy correction, $\Sigma$.
To the lowest order of PT the self--energy is given by
\be
\Sigma(v\cdot l)=-2iC_F\mu^{4-D}\int\frac{d^Dk}{(2\pi)^{D-1}}
\frac{\as v^2}{k^2(v\cdot l+v\cdot k)}\,.
\lab{se}
\ee
As was noticed in~\ci{Pol,IKR}, for $D=4$ the self--energy
$\Sigma(v\cdot l)$
contains a linear ultraviolet divergence which must be included
into the renormalization of the heavy quark masses,
$m_{b,c}\to m_{b,c}+\Delta m$. The last factor in~\re{W-IR} takes into account
the change of the plane wave of the heavy quarks,
$\exp(-i(m_bv -m_c v')\cdot x)$ with $x=-i\tt v'$,
due to renormalization of the heavy quark masses.
However, before the argument of the coupling constant is fixed,
the integral in~\re{se} contains only one scale and $\Sigma(v\cdot l)$
vanishes in the dimensional regularization as $(l\cdot v)\to 0$
leading to $\Delta m=0$. That is why self--energy corrections did not
appear in the one--loop expression~\re{WC1}.

Let us substitute $\as=\as(k_t)=\frac1{\beta_0}\int_0^\infty du
(k_t^2/\Lambda^2)^{-u}$ into \re{W-IR} and
\re{se} and perform the integration over gluon momenta to get
\be
W=\exp\lr{\frac{C_F}{2\pi\beta_0} \frac{(4\pi\mu^2/\Lambda^2)^\varepsilon}
                         {\Gamma(1-\varepsilon)}
\int_{\varepsilon}^\infty
du\ \lr{\Lambda\tt}^{2u}
\Gamma(-2u)\int_{-1}^1 dx\frac{(1-x^2)^{1-u}}
{(\coth\vartheta-x)^2}
+\tt\Delta m(1-\ch\vartheta)}
\,.
\lab{WC2}
\ee
To find $\Delta m$ we evaluate the self--energy correction \re{se}
\be
\Sigma(v\cdot l)= \Lambda\frac{C_F}{2\pi\beta_0}
\frac{(4\pi\mu^2/\Lambda^2)^\varepsilon}{\Gamma(1-\varepsilon)}
\int_{\varepsilon}^\infty du\
\Gamma^2(1-u)\Gamma(-1+2u)
(-2v\cdot l/\Lambda)^{1-2u}
\lab{mass}
\ee
and notice that the linear divergence of \re{se} corresponds
to the singularity of the integrand in \re{mass} at $u=1/2$.
Hence, $\Delta m$ is given by~\ci{BB}
\be
\Delta m = - \Lambda \frac{C_F}{2\beta_0}
\int \frac{du}{1-2u}\,,
\lab{mass1}
\ee
where integration is performed at vicinity of $u=1/2$.
Since $(\Lambda\tt)^{2u}=\e^{-u/\as(1/\tt)\beta_0}$,
the exponent in \re{WC2} has a form of the Borel transformation.
The integration in \re{WC2} is performed over all positive values of the
Borel parameters $u$ and the integrand contains the singularities
generated by the $\Gamma-$function at the points $u^*=\frac12,
1, ...$, the so--called infrared renormalons. Infrared
renormalons appear in \re{WC2} due to singularity of the coupling constant at
$k_t^2=\Lambda^2$. At the same time, integration in \re{WC2} over
small values of the Borel parameters, $u\ll\frac12$,
or equivalently over large momentum $k_t^2\gg\Lambda^2$ gives the result
for $W_{\rm pert}$ which after renormalization in the $\MS$ scheme
satisfies the RG equation \re{RG}.

The presence of the IR renormalon in the Wilson line implies, first, that
perturbative
expansion of $W$ is not well defined since we have to specify the
prescription for integrating IR renormalon singularities. Second, using
different prescriptions at $u^*=\frac{n}{2}$ we get the results
for the exponent of $W$ which differ in power corrections
$(\Lambda\tt)^n$ with $n=1,2...$. This means that perturbation
theory fails to describe power corrections to the Wilson line and
in order to make $W$ well defined one has to add nonperturbative
power corrections to the exponent of \re{WC2},
\begin{equation}
W(-i\tt,\vartheta)
=W_{\rm pert}(-i\tt,\vartheta) W_{\rm nonpert}(-i\tt,\vartheta)\,.
\lab{Wc}
\end{equation}
Thus, the form of nonperturbative power corrections to $W_{\rm nonpert}$
can be examined by calculating the contribution of the IR renormalon to
\re{WC2}. We start with the IR renormalon at $u^*=\frac12$
and expand the integrand in \re{WC2} near $u=u^*$ to find its
contribution to $W$ as
$$
\log W_{\rm pert}
\sim \tt\lr{\frac{\Lambda C_F}{2\beta_0}\int\frac{du}{1-2u}
+\Delta m} (1-\ch\vartheta) + \CO(\tt^2) \,.
$$
Using \re{mass1} we find that the contribution of the IR renormalon at
$u^*=\frac1{2}$ to the exponent of the Wilson line is compensated
by the contribution of the ultraviolet renormalon at
$u^*=\frac1{2}$ to $\Delta m$.
Hence, the leading renormalon contribution to the Wilson line is of order
$\CO(\Lambda^2\tt^2)$ and corresponds to $u^*=1$.
Thus, for the Wilson line \re{Wc} to be well defined, nonperturbative effects
should contribute to $W_{\rm nonpert}$ at the level of
${\cal O}(\Lambda^2\tt^2)$ power corrections. Expanding the integrand in
\re{WC2} near $u^*=1$ we find the corresponding contribution
as
\be
\log W_{\rm pert}\sim \tt\cdot 0 + \tt^2\Lambda^2\sh^2\vartheta\
\frac{C_F}{4\pi\beta_0}\int \frac{du}{1-u}
+\CO(\tt^3)\,.
\lab{exp}
\ee
We can ~parameterize nonperturbative power corrections to $W$
in terms of hadronic matrix elements by performing
expansion of the Wilson line $W$, defined in \re{W-def},
in powers of $\tt$:
$$
W_{\rm nonpert}(-it,\vartheta)=1 + a_1 \tt + \frac12 a_2 \tt^2 + \ldots\,,
$$
where $a_n=\bra{B} b_v^+ (iv'D)^n b_v \ket{B}$.
Using the general form of the matrix elements
$\bra{B} b_v^+ D_\mu b_v \ket{B}=c_1 v_\mu$ and
$\bra{B} b_v^+ D_\mu D_\nu b_v \ket{B}=c_2 g_{\mu\nu}+c_3 v_\mu v_\nu$
and taking into account the equation of motion, we get
$c_1=0$ and $c_2=-c_3=\mu_{0,\pi}^2/3$ with
$\mu_{0,\pi}^2=\bra{B} b_v^+ D^2 b_v \ket{B}$
being one of the fundamental parameters in HQET.
As a result, the small $\tt$ expansion of $W$ is given by
\begin{equation}
W_{\rm nonpert}(-i\tt,\vartheta)
=1 + 0\cdot \tt + \frac16 \tt^2 \mu_{0,\pi}^2 \sh^2\vartheta
+{\cal O}(\tt^3)\,.
\lab{nonpert}
\end{equation}
Comparing this expression with \re{exp} we find an agreement with
the IR renormalon analysis. Namely, the coefficient in front of
$\tt$ vanishes
and the term with $\tt^2$ has the same dependence on the angle $\vartheta$.
Moreover, perturbative $\CO(t^2)$ corrections to $W_{\rm pert}$
contain infrared renormalon ambiguity while nonperturbative $\CO(t^2)$
corrections to $W_{\rm nonpert}$ have ultraviolet renormalon ambiguity
in the definition of the matrix element $\mu_{0,\pi}^2$. However, the sum
of both corrections becomes unique due to cancellation of IR and
UV renormalons and it defines the nonperturbative parameter $\mu_{\pi}^2$.

Being taken separately,
$W_{\rm pert}$ and $W_{\rm nonpert}$ suffer from the presence of the
IR and UV renormalons, respectively. The contributions of
IR and UV renormalons cancel each other in the product \re{Wc} to
make the Wilson line as well as nonperturbative parameters like
$\mu_\pi^2$ well defined. The fact, that the IR renormalons
appear in the exponent of $W_{\rm pert}$ implies that the UV
renormalons should exponentiate in $W_{\rm nonpert}$ as well and using
the short distance expansion \re{nonpert} one can choose the ``minimal''
anzatz for nonperturbative part $W_{\rm nonpert}(-it,\vartheta)$:
\be
W_{\rm nonpert}(-i\tt,\vartheta)
=\exp\lr{\frac16 \tt^2 \mu_\pi^2 \sh^2\vartheta
+{\cal O}(\tt^3)}\,,
\lab{ans}
\ee
where the expansion in the exponent does not contain a term linear
in $\tt$. Substituting \re{ans} into the definition \re{fin} of the structure
function and performing Fourier integration, we obtain the nonperturbative
anzatz for the structure function
\be
F_{\rm nonpert}(\varepsilon,\vartheta)=\frac1{\sqrt{2\pi\sigma^2}}
\exp\lr{-\frac{(\varepsilon-\varepsilon_0 \ch\vartheta)^2}
       {2\sigma^2}
       }\,, \qqquad
\sigma^2=\frac13 {\mu_\pi^2\sh^2\vartheta}\,,
\lab{toy}
\ee
which describes a Gaussian distribution of the residual energy $\varepsilon$
with the width $\sigma$ depending on the angle $\vartheta$.
For $\vartheta\to 0$
this function becomes $\delta(\varepsilon-\varepsilon_0)$. This means
that the light components of the $\B$ meson do not interact with outgoing
heavy quark as it should be due to the conservation of the heavy quark
current at zero recoil.

\subsection{Renormalized sum rules for heavy--heavy transitions}

Finally, after substitution of \re{nonpert} and \re{pert} into
\re{Wc} and \re{W-F} we arrive at the sum rule
\ba
\int_{\varepsilon_0}^\infty d\varepsilon\
\e^{-\tt(\varepsilon-\varepsilon_0)}
F(\vartheta,\varepsilon)&=&\exp\lr{-\tt\varepsilon_0(\ch\vartheta-1)
-2\int_{2/\tt}^\mu \frac{dk_t}{k_t} \Gamma_{\rm cusp}(\vartheta,\as(k_t))
+ \gamma(\vartheta,\as(2/\tt))}
\nonumber\\
&&\times
\left(1+\frac16\mu_\pi^2\tt^2\sh^2\vartheta+{\cal O}(\tt^3)\right).
\lab{Bj0}
\ea
These sum rule take into account
perturbative and nonperturbative effects and generalize the Bjorken sum
rule~\ci{Bj}. With eq.\re{SFFF} taken into account,
the l.h.s.\ of~\re{Bj0} is given by infinite sum over excited states $n$.
For the lower states $n$ the expansion in \re{SFFF} is well defined
and after integration in \re{Bj0} it leads to the power series in $\tt$
similar to that in \re{nonpert}.  But for higher states $n$, corresponding
to large residual energy of light particles in the final state $X_c$, the
sum in \re{SFFF} is determined by hard gluon emission and
after integration in \re{Bj0} it gets $\log\tt$ corrections similar to
\re{pert}. The exponential factor in the l.h.s.\ of \re{Bj0} implements
the smooth exponential cut off on the residual energy at
$\varepsilon-\varepsilon_0\sim1/\tt$. This should be compared with
a sharp cut off $\theta(\varepsilon-1/\tt)$ used in~\ci{IW2,Vol}.

We stress that the sum rules contain two completely independent
scales, $\mu$ and $\tt$, and the both sides of~(\ref{Bj0}) have the
same dependence on $\mu$ and separately on $\tt$. The inclusive
structure function and the exclusive form factors do not depend on
$\tt$ while their dependence on the normalization point $\mu$
is described by the RG equations \re{RG} and \re{RG1}. We notice that the
exponent of the integral of the cusp anomalous dimension can be combined
with the structure function to produce $F(\vartheta,\varepsilon)$
at $\mu=2/\tt$. Therefore the sum rule \re{Bj0} can be written in the
simpler form
\begin{equation}
\int_{\varepsilon_0}^\infty d\varepsilon\
\e^{-\tt(\varepsilon-\varepsilon_0)}
F(\vartheta,\varepsilon)\bigg |_{\mu=2/\tt}
=\e^{-\tt\varepsilon_0(\ch\vartheta-1)+\gamma(\vartheta,\as(2/\tt))}
\left(1+\frac16\mu_\pi^2\tt^2\sh^2\vartheta+{\cal O}(\tt^3)\right).
\lab{Bj}
\end{equation}
which contains only one scale $\tt$.
In these sum rules $\tt$ is a free parameter and
by differentiating the both sides of \re{Bj} with respect to $\tt$,
we can get generalizations of the Voloshin~\ci{Vol}
and the ``third''~\ci{BGSUV} sum rules.
We notice however that the possible values of $\tt$ in the sum rules
are restricted as
\be
1/m_c \ll \tt \ll 1/\Lambda\,.
\lab{cond}
\ee
The upper limit for $\tt$ follows from the condition for
the expansion in the r.h.s.\ of \re{nonpert} and \re{Bj}
to be well defined. To understand
the lower limit we recall that the structure function
$F(\vartheta,\varepsilon)$ was defined before under the additional
condition that the $\c$ quark should be ``heavy'', or equivalently
the residual energy of the final states $X_c$ has to be small,
$\varepsilon \ll m_c$. This leads to $\tt \gg 1/m_c$ because the
residual energy in \re{Bj} is cut at $\varepsilon-\varepsilon_0
\sim 1/\tt$ with $\varepsilon_0 \sim \Lambda$.

Considerable simplifications occur in the sum rules in the small velocity
limit, $\vartheta\to 0$. Let us expand the both sides of \re{SFFF}
to the first order in $\vartheta^2$. The ground state form factor is%
{}~\ci{IW}
$$
\xi^{(0)}(\ch\vartheta)=1-\rho^2(\mu)\ (\ch\vartheta-1)+\ldots
$$
where $\rho^2(\mu)$ is the slope parameter. The transition form factors to
higher  S--wave states behave as
$\xi^{(n)}(\ch\vartheta)={\xi^{(n)}}'(1)(\ch\vartheta-1)+\ldots$
and may be ignored in \re{SFFF}. Contributions of the transition
form factors to the P--wave states, $\tau_{1/2}$
and $\tau_{3/2}$, have the threshold suppression $\ch\vartheta-1$ in%
{}~\re{SFFF},
and, to $\CO(\vartheta^2)$ order, the form factors may be taken at
$\vartheta=0$. Moreover, contributions of the transitions
to D--wave (and higher) states indicated by dots in~\re{SFFF} have the
threshold suppression $(\ch\vartheta-1)^2$ (and stronger), and may be ignored.
The anomalous dimension \re{gamma} behaves in the small $\vartheta$ limit
as
$$
\gamma(\vartheta,\as)=\vartheta^2\gamma_0(\as)+{\cal O}(\vartheta^4)\,,
\qquad \gamma_0(\as)=-\frac59 C_F \alpi+\CO(\as^2)\,.
$$
Expanding~(\ref{Bj0}) to the $\vartheta^2$ order, we obtain
the generalization of the Bjorken sum rule~\ci{IW2}
\begin{equation}
\sum_n|\tau_{1/2}^{(n)}(1)|^2 e^{-\delta_{1/2}^{(n)}\tt}
+ 2\sum_n|\tau_{3/2}^{(n)}(1)|^2 e^{-\delta_{3/2}^{(n)}\tt}
= -\frac14+ \rho^2(2/\tt) + \gamma_0(\as(2/\tt)) - \frac12\varepsilon_0\tt
+ \frac16\mu_\pi^2\tt^2 + {\cal O}(\tt^3)
\lab{Bj2}
\end{equation}
where $\delta_{1/2(3/2)}^{(n)}=\varepsilon_{1/2^+(3/2^+)}^{(n)}-\varepsilon_0$
are the excitation energies of the P--wave states.
Here, the values of form factors $\tau_{1/2}(1)$ and $\tau_{3/2}(1)$
do not depend on the normalization point $\mu$ while the slope parameter
$\rho$ is defined at $\mu=2/\tt$.

Differentiating~\re{Bj2} in $\tt$, we can obtain the Voloshin~\ci{Vol}
and the ``third''~\ci{BGSUV} sum rules:
\ba
&&\sum_n|\tau_{1/2}^{(n)}(1)|^2 \delta_{1/2}^{(n)}
\e^{-\delta_{1/2}^{(n)}\tt}
+ 2\sum_n|\tau_{3/2}^{(n)}(1)|^2 \delta_{3/2}^{(n)}
\e^{-\delta_{3/2}^{(n)}\tt}
\nonumber\\
&&\qqqquad\qqqquad\qqqquad= \frac12\varepsilon_0
- \frac{d}{d\tt}\left(\rho^2(2/\tt)+\gamma_0(\as(2/\tt))\right)
- \frac13\mu_\pi^2\tt + {\cal O}(\tt^2),
\lab{Vol}
\\[3mm]
&&\sum_n|\tau_{1/2}^{(n)}(1)|^2 \left(\delta_{1/2}^{(n)}\right)^2
\e^{-\delta_{1/2}^{(n)}\tt}
+ 2\sum_n|\tau_{3/2}^{(n)}(1)|^2 \left(\delta_{3/2}^{(n)}\right)^2
\e^{-\delta_{3/2}^{(n)}\tt}
\nonumber\\
&&\qqqquad\qqqquad\qqqquad= \frac13\mu_\pi^2
+ \frac{d^2}{d\tt^2}\left(\rho^2(2/\tt)+\gamma_0(\as(2/\tt))\right)
+ {\cal O}(\tt)\,,
\nonumber
\ea
where an arbitrary cut-off parameter $t$ satisfies condition \re{cond}.

\section{Heavy to light transitions}
\lab{HL}

Let us consider now the decay of the $\B$ meson in the limit of the
``light'' $\c$ quark mass, $m_c^2\ll\Lambda m_b$. As was stressed in
sect.~1, for small invariant mass of the final state,
$m_X^2-m_c^2=\CO(m_c\Lambda)$, the $\c$ quark behaves effectively as
a heavy quark and one can apply the results of the previous section
to describe the $\B\to X_c+W$ decay in terms of the structure function
of heavy to heavy meson transition. As the mass of the final state
increases, $m_b\Lambda \gg m_X^2\gg m_c^2$, we encounter a new
situation in which the energy of the light components of $X_c$ is
much bigger $m_c$ and the $\c$ quark becomes effectively
massless. As a consequence, the heavy quark expansion in powers
of $1/m_c$ is not applicable in this case and the decomposition of the
momentum $p_X$ of the final state as well as the notion of
the residual energy of the final state, $\varepsilon_X$, do not have
a sense. In what follows, we neglect the mass of the $\c$ quark
and analyze the differential rate of the decay $\B\to X_c+W$ in the
leading $1/m_b$ limit.

In the rest frame of the $\B$ meson the $\b$ quark decays into the $\c$
quark which has a large energy $m_b-W_0=\CO(m_b)$ and small
invariant mass $(m_bv-W)^2\sim m_X^2$. The total momentum of the final state,
$p_X=m_Bv-W$, is equal to the sum of momenta of the $\c$ quark and
light components of the $\B$ meson. It is convenient to introduce the
light--cone variables $p^\pm =\frac1{\sqrt2}(p^0\pm p^3)$ and
${\bf p}_t=(p^1,p^2)$ for an arbitrary momentum $p$.%
\footnote{We recall that in the light--cone variables,
          for arbitrary 4--vectors $p$ and $k$,
          we have $p\cdot k=p^+k^-+p^-k^+-{\bf p}\cdot{\bf k}$.}
Then, in the rest frame
the $\B$ meson we have $p_B=\frac{m_B}{\sqrt2}(1^+,1^-,{\bf 0})$
and $W=(W^+,W^-,{\bf 0})$. Particles in the final states move close to
the light--like direction which we choose to be the ``$-$'' direction on the
light--cone,
$$
p_X^- = \CO(m_b)\,,\qquad
p_X^+ = \CO(\Lambda)
$$
with $m_X^2=2p_X^+ p_X^- =\CO(m_b\Lambda)$. Propagating into the final
state, the $\c$ quark can produce a jet of energetic collinear particles
which move in the ``$-$'' direction and are accompanied by soft radiation
including the light components of the $\B$ meson~\ci{KS}.
This should be compared with the case of heavy to heavy
transition in which the residual energy $\varepsilon$ is too small with
respect to $m_c$ for the outgoing heavy quark to create a jet. Thus,
in the case of ``light'' $\c$ quark we have to take into account a new
possibility for the $\c$ quark to create energetic jet of particles. This
makes the analysis more complicated compared to the previous case.

\subsection{IR factorization}

The momentum $p_X$ of the final state is distributed among collinear
and soft particles which interact with each other. However since the
``$-$'' component of the momentum of collinear particles is much bigger
the ``$-$'' component of the momentum of soft particles, this interaction
depends only on the ``$+$'' components of the corresponding momenta which
are of the same order for soft and collinear particles.
The total momentum of the final state, $p_X$, is equal to the
sum of momentum of the light component of the $\B$ meson, $\varepsilon_0 v$,
momentum $k$ of soft gluons emitted by the $\b$ and $\c$ quarks in
the partonic subprocess and momentum $l$ of collinear particles.
Then, the phase space for $k^+$ component is
$$
-\varepsilon_0 v^+ < k^+ < p_X^+\,,
$$
where the lower limit follows from condition for the total momentum
of soft radiation, $\varepsilon_0 v+k$, to be time--like, the upper limit
corresponds to the extreme case when the total momentum of the final state
$p_X^+=m_B v^+ - W^+=\CO(\Lambda)$ is formed by soft particles while collinear
particles have zero ``$+$'' momentum, $l^+=0$.

Performing the analysis of
the Feynman diagrams corresponding to the differential rate of the decay,
we find that the contributions of jet and soft subprocesses to
$d\Gamma(\B\to X_c+W)$ may be factorized in the leading $1/m_b$ limit into
the form~\ci{KS}
\be
\frac{d^2\Gamma}{d W^2 d W_0}
=\int_{-\varepsilon_0}^{p_X^+/v^+} d\varepsilon\, f(\varepsilon)\,
\sigma_0(p_X^+/v^+-\varepsilon)
\lab{DIS}
\ee
where the structure function $f(\varepsilon)$ takes into account interaction
of the $\b$ and $\c$ quarks with soft radiation while
$\sigma_0(p_X^+/v^+-\varepsilon)$ describes interaction of the $\c$ quark and
collinear particles in the final state with momentum
$l^+=p_X^+-\varepsilon v^+$.

The function $f(\varepsilon)$ is analogous to the structure function of the
heavy to heavy transition but in contrast with the previous case the velocity
$v'$ of the outgoing heavy quark should be replaced by ``$-$''
light--cone direction in which massless $\c$ quark moves into the final state.
Then, replacing $v'$ by the light--like vector $y^\mu=(0^+,y^-,{\bf 0})$
in the definition~\re{fin} we find the contribution of the
soft subprocess to the differential rate as a Fourier transformed
Wilson line expectation value~\ci{Kor}
\be
f(\varepsilon) = v^+ \int_{-\infty}^\infty \frac{dy^-}{2\pi}\,
\e^{i\varepsilon v^+y^-}\,W_C(v^+y^-)\,,
\lab{f}
\ee
with $W_C$ defined in~\re{W-def} for an arbitrary path $C$.
Here, the integration path $C$ goes from $-\infty$ to point $0$ along
the $\b$ quark velocity $v$, then along the light--cone ``$-$''
direction to point $y$, and then from $y$ to $-\infty$ along $-v$.
The structure function $f(\varepsilon)$ is defined in the limit $m_b\to\infty$,
and its properties don't depend on $m_b$. The only place where $m_b$
dependence appears is the factorization formula~\re{DIS}.

Comparing the definition of the structure functions of heavy--heavy
and heavy-light transitions, eqs.\re{fin} and \re{f}, respectively,
we find that in both cases the structure function is given by
a one--dimensional Fourier transformed Wilson line expectation value.
The Fourier integration is performed along the direction of the outgoing
$\c$ quark which coincides with a time--like vector $v'$ for heavy--heavy
transition and with the light--cone ``$-$'' direction of the jet for
heavy--light transition. The argument of the Fourier transformation
is equal in both cases to the projection of the total momentum
of the soft radiation onto the $\c$ quark direction.
In the heavy--heavy transition it is equal to $v'\cdot(k'-k)$
with the residual momenta $k$ and $k'$ defined in~\re{dec}
and~\re{res}, while in the heavy--light case it is $k^+=\varepsilon v^+$.
Since in the heavy--light transition the final state $X_c$ consists of the
jet and the soft radiation, the momentum $k$ is given by
$k=p_X-\varepsilon_0 v-l$.

Similar to the structure function $F$,
the function $f$ depends only on the properties of the $\B$ state and
not on particular short distance subprocess. This function
describes the distribution of the light fields' momenta $k^+=\varepsilon v^+$
in the $\B$ meson, or equivalently the probability to find the $\b$ quark
with certain light--cone residual momentum projection $-k^+$ inside the $\B$
meson. The factorized expression~\re{DIS} for the differential rate has a
simple partonic interpretation with $f(\varepsilon)$ being the distribution
function of the $\b$ quark inside the $\B$ meson and $\sigma_0$ the partonic
subprocess. In contrast with the distribution function, the partonic subprocess
$\sigma_0$ can be calculated perturbatively and it gets large
perturbative corrections to all order of PT which can be resummed~\ci{KS}.

\subsection{Evolution equation for the structure function}

The heavy quark distribution function has a nonperturbative origin
and as a function of the factorization scale $\mu$ it obeys the
GLAP--type evolution equation~\ci{Kor}. Comparing the definitions
of the structure functions of heavy--heavy and heavy--light transitions,
\re{fin} and~\re{f} respectively,
we notice that there is an essential difference in the definition of the
integration path entering the Wilson line.
Namely, the time--like direction $v'$ in~\re{fin} is replaced by the
light--cone ``$-$'' direction in~\re{f}. This suggests to apply
the results of the section~2.3 by replacing $v'_\mu$ by
the light--like vector $y^\mu=(0^+,y^-,{\bf 0})$. In particular,
the angle $\vartheta$ between the vectors $v$ and $v'$ should be
replaced by the angle between $v$ and $y$ which is equal to
infinity for $y^2=0$. However, trying to perform the limit
$\vartheta\to\infty$ in~\re{RG} we notice that the cusp anomalous dimension
has the following asymptotics~\ci{KR} in the large $\vartheta$ limit
\be
\Gamma_{\rm cusp}(\as,\vartheta)
\stackrel{\vartheta\to\infty}{=}\vartheta \Gamma_{\rm cusp}(\as)
+\CO(\vartheta^0)\,,\quad
\Gamma_{\rm cusp}(\as)=\alpi C_F+\left(\alpi\right)^2
C_F\left(C_A\left(\frac{67}{36}-\frac{\pi^2}{12}\right)-N_f\frac5{18}\right)
\lab{as}
\ee
and the RG equation~\re{RG} becomes meaningless as $\vartheta\to\infty$.
This explains why the structure function of the heavy--light transition does
not satisfy the RG equation~\re{RG}. For $y^2=0$ the Wilson line
entering into~\re{f} has additional light--cone singularities which
modify the renormalization properties of the Wilson line~\ci{Kor,KK,KM}.

To find the RG equation for the light--like Wilson line we perform
the one--loop calculation of $W_C$ using~\re{WC1} and replacing $v'$ by $y$.
For $y^2=0$, the result of integration contains a double pole in
$\epsilon$ and after renormalization in the $\MS$--scheme we obtain
the one--loop expression for the light--like Wilson line as~\ci{Kor}
\be
W_{\rm pert}(v^+y^-)=1+\alpi C_F\left(-L^2 + L - \frac5{24}\pi^2\right),
\qquad L=\log[i(\mu v^+y^- - i0)] + \gamma_E
\,,
\lab{one1}
\ee
where ``$-i0$'' prescription comes from the position of the pole
in the free gluon propagator in the coordinate representation~\ci{Kor},
and $\gamma_E$ is the Euler constant. It is convinient to absorb the
Euler constant into the definition of the renormalization parameter $\mu$.
Due to additional light--cone singularities
the Wilson line has a double logarithm of $\mu$ to one--loop order.
The renormalization properties of the light--like Wilson lines have been
studied in~\ci{Kor,KK,KM}. It was shown that the Wilson line $W_C$ satisfies
the following RG equation
\be
\left(\mu\frac{\partial}{\partial\mu}+\beta(g)\frac{\partial}{\partial g}
+\Gamma_{\rm cusp}(\as)[\log(\mu v^+y^--i0)+\log(-\mu v^+y^-+i0)]
 +\Gamma(\as)\right) W(v^+y^-)
 =0
\lab{RG2}
\ee
with $\Gamma_{\rm cusp}(\as)$ defined in~\re{as} and $\Gamma(\as)$ some
anomalous dimension. Two--loop expression for $\Gamma(\as)$ is given by~\ci{KM}
$$
\Gamma(\as)=-\alpi C_F
-\left(\alpi\right)^2C_F\left[
\left(\frac7{18}\pi^2+\frac{37}{108}-\frac94\zeta(3)\right)C_A
 -\left(\frac1{54}-\frac1{18}\pi^2\right) N_f
\right]
$$
One easily verifies that the one--loop Wilson line~\re{one1} satisfies the
RG equation~\re{RG2}. In perturbation theory the light--like Wilson line
is a dimensionless function of $\mu v^+y^-$. Solving~\re{RG2}
we find the perturbative expression for $W$:
\be
W_{\rm pert}(-i\tt)=
\exp\lr{-\int_{1/\tt}^\mu \frac{dk_t}{k_t}
\lr{2\Gamma_{\rm cusp}(\as(k_t)) \log({k_t}{\tt})+\Gamma(\as(k_t))}
+\gamma(\as(1/\tt))}\,.
\lab{sol}
\ee
The one--loop expression for the
integration constant $\gamma(\as)$ can be found from comparison
of~\re{sol} with~\re{one1} as
$$
\gamma(\as)=-\alpi C_F \frac{5}{24}\pi^2+\CO(\as^2)\,.
$$
Expression~\re{sol} should be compared
with analogous expression~\re{pert} for heavy--heavy transition.

To find the dependence of the structure function on the
renormalization point we use the definition~\re{f} and perform the Fourier
transformation of the both sides of the RG equation~\re{RG2} with respect
to $y^-$. Since the anomalous dimension in~\re{RG2} depends on $y^-$,
the resulting equation for the structure function is the evolution
equation~\ci{Kor,KM} rather than the RG type equation:
\be
\left(\mu\frac{\partial}{\partial\mu}+\beta(g)\frac{\partial}{\partial g}
\right) f(\varepsilon) = \int_{-\varepsilon_0}^{\varepsilon}
d\varepsilon'\, P(\varepsilon-\varepsilon')\, f(\varepsilon').
\lab{GLAP}
\ee
Here, the evolution kernel is given by
\be
P(\varepsilon)=-\int_{-\infty}^{\infty} \frac{dt}{2\pi}\,
\e^{i\varepsilon t}\,\left(2\Gamma_{\rm cusp}(\as)\log(\mu t-i0)+\Gamma(\as)
\right)
\lab{P}
\ee
and the ``$-i0$'' prescription leads to the spectral property
$$
P(\varepsilon)=0 \qquad \mbox{for $\varepsilon<0$}\,.
$$
This is the reason for setting to $\varepsilon$ the upper limit for
$\varepsilon'$ in~\re{GLAP}. The kernel $P(\varepsilon)$ has the dimensionality
1/energy, and depends on the dimensional argument $\varepsilon$. Therefore it
can only contain two terms $\theta(\varepsilon)/\varepsilon$ and
$\delta(\varepsilon)$. After integration in~\re{P} we  obtain \ci{Kor,KM}
\be
P(\varepsilon)=-2\Gamma_{\rm cusp}(\as)
\lr{\frac{\theta(\varepsilon)}{\varepsilon}}_+^{(m)}
+\lr{2\Gamma_{\rm cusp}(\as)\log\frac{m}{\mu}+\gamma_E-\Gamma(\as)}
\delta(\varepsilon)\,,
\lab{P1}
\ee
where the distribution $(\theta(\varepsilon)/\varepsilon)_+^{(m)}$ is defined,
for an arbitrary mass scale $m$, by the formula
$$
\int_{-\infty}^\infty d\varepsilon\,
\lr{\frac{\theta(\varepsilon)}{\varepsilon}}_+^{(m)}\,\varphi(\varepsilon)
=\int_0^m d\varepsilon\,\frac{\varphi(\varepsilon)-\varphi(0)}{\varepsilon}
+\int_m^\infty d\varepsilon\,\frac{\varphi(\varepsilon)}{\varepsilon}
$$
for any smooth test function $\varphi(\varepsilon)$ which falls quickly enough
for the last integral to converge. We have
$(\theta(\varepsilon)/\varepsilon)_+^{(m')}
-(\theta(\varepsilon)/\varepsilon)_+^{(m)}=-\delta(\varepsilon)\log(m'/m)$,
therefore the kernel~\re{P1} does not, in fact, depend on $m$.
We notice that the evolution kernel contains terms proportional to the
$\delta(\varepsilon)$ function which depend on the renormalization scale $\mu$.
In the solution of the evolution equation~\re{GLAP}
for the distribution function $f(\varepsilon)$, these terms can be
absorbed into $\varepsilon$--independent factor. Moreover, in the factorized
expression~\re{DIS} for the
differential rate the $\mu$ dependence of the structure function
is compensated by the $\mu$ dependence of the partonic subprocess $\sigma_0$.

It is important to realize that the renormalization properties of the
distribution function \re{GLAP} follow from the
properties of the Wilson lines. In particular, the definition of the
Wilson line in~\re{W-def} does not involve time--ordering of gauge field.
Had we start with the same definition~\re{f} of the distribution
function but with the time--ordering included, the resulting function
would satisfy the Brodsky--Lepage evolution equation~\ci{KK} for the
meson wave function rather than the GLAP equation.

The heavy quark distribution function $f(\varepsilon)$ defined in \re{f} is
closely related to the distribution function of the $\b$ quark inside
the $\B$ meson, $f_{\b/\B}(x)$, \ci{Kor,KM}. Namely, in the leading heavy
quark limit both function coincide in the end-point region $x\sim 1$
provided that $x=1-\varepsilon/m_b$. As a consequence, the evolution
kernel
%The evolution kernel~\re{P1} satisfies the following equation
%$$
%\varepsilon P(\varepsilon) = -2\Gamma_{\rm cusp}(\as)\,\theta(\varepsilon)\,.
%$$
%The kernel
$P(\varepsilon)$ coincides with the standard GLAP quark splitting
function $P_{qq}(z=1-\varepsilon/m_b)$ in the limit $z\to1$~\ci{Kor,KM}.
However, there is an important difference between these two
distribution functions
due to the fact that they have different
spectral properties. In contrast to the ordinary GLAP equation, the
kernel~\re{P1} does not
allow one to obtain the anomalous dimensions of the HQET local operators
$O_n=b_v^\dagger(iD^+)^n b_v$. The moments of the structure function
$\int_{-\varepsilon_0}^\infty d\varepsilon \varepsilon^n f(\varepsilon)$,
after renormalization of $f(\varepsilon)$, still contain ultraviolet
divergences, because $\varepsilon$ (having the dimensionality of energy)
varies up to infinity.
%The divergences of the matrix elements of the local
%operators $O_n$ are not exhausted by the divergences of the $d{\bf k}_t$
%integrations which determine renormalization of the structure function.
This sharply differs from the ordinary QCD case, where integrals in
the dimensionless variable $x$ varying from 0 to 1 due to spectral
properties of $f_{\b/\B}(x)$ and can never contain
ultraviolet divergences.

\subsection{IR renormalons in the light--like Wilson line}

The general solution to the RG equation~\re{RG2} has a form similar
to~\re{Wc}
\be
W(v^+y^-)=W_{\rm pert}(v^+y^-) W_{\rm nonpert}(v^+y^-)
\lab{Wc1}
\ee
with $W_{\rm nonpert}(v^+y^-)$ being a nonperturbative boundary value
for the Wilson line. To find the properties of $W_{\rm nonpert}(v^+y^-)$,
we perform the analysis of the IR renormalons in the light--like Wilson line.
Using~\re{W-IR} and replacing $v'$ by the light--like vector $y$
we obtain
\ba
W_{\rm pert}(v^+y^-)&=&\exp\lr{
            2C_F\mu^{4-D}\int\frac{d^Dk}{(2\pi)^{D-2}}\,
            \as\delta(k^2)\theta(k_0)
            \lr{\frac{y}{(yk)}-\frac{v}{(vk)}}^2
            \lr{1-\e^{-i(yk)}} +\CO(\as^2)
            }
\nonumber \\
   &\times&\exp\lr{-i\Delta m v\cdot y}
\,,
\lab{WC11}
\ea
where $\Delta m$ comes from the self--energy corrections to the $v$--line,
eqs.~\re{se} and~\re{mass1},
while the light--cone $y$--line is protected from these corrections.
Integration over gluon momentum leads to the expression for the light--like
Wilson line
\be
W_{\rm pert}(-i\tt)
=\exp\lr{\frac{C_F}{\beta_0} \frac{(4\pi\mu^2/\Lambda^2)^\varepsilon}
                         {\Gamma(1-\varepsilon)}
\int_{\varepsilon}^\infty
du\,\lr{\Lambda\tt}^{2u}
\frac{\Gamma(-2u)
(1-u)}{
\sin(\pi u)}
}
\exp\lr{-\Delta m \tt}
\,.
\lab{WC21}
\ee
which contains the IR renormalons at $u^*=\frac12,1,\ldots$\ .
Different choices of the
prescription in~\re{WC21} lead to results which differ in power corrections
of the form $(\Lambda\tt)^{2u^*}$. The leading power
corrections correspond to the left--most singularity at
$u^*={1\over 2}$. We find however that similarly to the
case of the heavy--heavy transition the residue of the
first exponent of~\re{WC21} at $u={1\over 2}$ is compensated
by the residue of $\Delta m$, eq.\re{mass1}, at $u={1\over 2}$
$$
\log W_{\rm pert}(-i\tt)=
-\lr{\frac{\Lambda C_F}{2\beta_0}\int \frac{du}{1-2u}+\Delta m}\tt
+\CO(\tt^2)
=0\cdot \tt+\CO(\tt^2)\,.
$$
Thus, for $u^*={1\over 2}$
the contributions of infrared and ultraviolet
renormalons to the light--like Wilson line cancel each other, so that the
leading IR renormalon appears in~\re{WC21} at $u^*=1$
$$
\log W_{\rm pert}(-i\tt)=
0\cdot\tt+\tt^2\Lambda^2\frac{C_F}{4\pi\beta_0}\int \frac{du}{1-u}
+\CO(\tt^3)\,.
$$
We conclude~\ci{KS}
that resummed perturbation theory generates (but fails to
describe uniquely) $\CO(\Lambda^2\tt^2)$ power corrections to the Wilson
line $W_C$. At the same time, this means that nonperturbative effects should
also contribute to $\CO(\Lambda^2 t^2)$ power corrections to make the
Wilson line~\re{Wc1} well defined. Indeed, performing the small $t$
expansion of the Wilson line we find the expression similar to~\re{nonpert}
\be
W_{\rm nonpert}(-i\tt)=1 + 0\cdot\tt + \frac{\mu_\pi^2}6 \tt^2
\lab{non}
\ee
in which the coefficient in front of $t$ vanishes due to the equation of
motion for the heavy quark. Repeating the arguments in favor of~\re{ans}
we choose the nonperturbative anzatz for the light--like Wilson line as
\be
W_{\rm nonpert}(-i\tt)=\exp\lr{\frac{\mu_\pi^2}6 \tt^2 + \CO(\tt^3)}\,.
\lab{ans1}
\ee
Then, after its substitution into~\re{f} we find the nonperturbative
anzatz for the heavy quark distribution function~\ci{KS}
\be
f_{\rm nonpert}(\varepsilon)=\frac1{\sqrt{2\pi\sigma^2}}
\exp\lr{-\frac{\varepsilon^2}{2\sigma^2}}\,,
\qquad \sigma^2=\frac13 \mu_\pi^2\,,
\lab{toy1}
\ee
which describes a Gaussian distribution around $\varepsilon=0$
with the width $\sigma$. Since we neglected $\CO(t^3)$ and higher corrections
in the exponent~\re{ans1}, this function does not vanish outside the physical
region $\varepsilon>-\varepsilon_0$ and it is valid only at the vicinity of
$\varepsilon=0$.

\subsection{Renormalized sum rules for heavy--light transitions}

To find the sum rules for the heavy quark distribution function
we invert the relation~\re{f} and express the Wilson line in
terms of $f(\varepsilon)$,
$$
W_C(-i\tt)=\int_{-\varepsilon_0}^{\infty} d\varepsilon\,\e^{-\varepsilon t}\,
f(\varepsilon).
$$
Now we substitute the expressions~\re{Wc1}, \re{sol}, and~\re{non}
for the Wilson line to get
\baa
\int_{-\varepsilon_0}^{\infty} d\varepsilon\,\e^{-\varepsilon t}\,
f(\varepsilon)&=&\exp\lr{-\int_{1/\tt}^\mu \frac{dk_t}{k_t}
\lr{2\Gamma_{\rm cusp}(\as(k_t))\log({k_t}{\tt})+\Gamma(\as(k_t))}
+\gamma(\as(1/\tt))}\\
&\times&\lr{1 + \frac{\mu_\pi^2}6 \tt^2 +\CO(\tt^3)}
\eaa
Using the evolution equation~\re{RG2} we check that the both sides of the sum
rules have the same dependence on the renormalization scale $\mu$. Putting
$\mu=1/\tt$ we find the renormalized Bjorken sum rule
\be
\int_{-\varepsilon_0}^{\infty} d\varepsilon\,\e^{-\varepsilon t}\,
f(\varepsilon){\bigg|_{\mu=1/\tt}}=
\e^{\gamma(\as(1/\tt))}\lr{1 + \frac{\mu_\pi^2}6 \tt^2 +\CO(\tt^3)}
\lab{Bj1}
\ee
in which the distribution function is defined at the scale $\mu=1/\tt$.
Here, $\tt$ is a free parameter which implements smooth exponential
cutoff on large values of the light--cone momentum component
$k^+=\varepsilon v^+$. Differentiating~(\ref{Bj1}) in $\tt$, we obtain
Voloshin and the ``third'' sum rules.

\section{Conclusion}
\lab{Concl}

In this paper, we have considered properties of the structure functions
of inclusive heavy meson decays $\B\to W+X_c$ in two extreme cases
of heavy and light $\c$ quark corresponding to heavy--heavy and
heavy--light meson transitions respectively.
They are given by Fourier transforms%
{}~\re{fin} and \re{f}
of the diagonal matrix elements of the Wilson line going from $-\infty$
to $0$ along the heavy meson velocity $v$, then to some point $y$, and
then back to $-\infty$ along $-v$. In the case of heavy--heavy transitions,
$y=v't$ is time--like, while in the case of heavy--light transitions
$y$ is light--like: $y^2=0$. As a function of the factorization scale,
the former structure function is
multiplicatively renormalizable~(\ref{RG}), with the anomalous dimension
being twice cusp anomalous dimension of Wilson lines; the later one obeys the
GLAP--type evolution equation~(\ref{GLAP}) with the kernel~(\ref{P1}).

Several lowest moments of the structure functions are given by simple
sum rules, the Bjorken~\cite{Bj,IW2}, Voloshin~\cite{Vol,Bur} and
``third''~\cite{BGSUV} ones. However, being taken in their original
form these sum rules turn out to be divergent due to contribution
of the states $X_c$ with large residual energy.
Since contribution of these states is essentially perturbative, it
should be included to the coefficient functions, and should be
therefore excluded from the integrals of the matrix elements given by
the sum rules. This means that had we perform consistent factorization
procedure, the integrals over residual energy of the states $X_c$ in the
sum rules should be cut off in some way at the factorization scale
$\mu$~\cite{IW2,Vol,Bur,BGSUV}. This is also necessary in order to
reproduce the correct $\mu$ dependence of parameters entering the
theoretical side of the sum rules. However, no consistent method to
implement such a cut--off was proposed.

In this paper, we introduced a soft exponential cut--off on the
residual energy of the states $X_c$ instead of a sharp step--function
\ci{IW2,Vol}. We show that with such a cut--off included, the sum rules
involve the Wilson lines analytically continued to
an imaginary separation $t$ or $y^-$. At a sufficiently
large factorization scale $\mu$, the corresponding Wilson lines can be
calculated theoretically, as a perturbative expansion plus a series
on nonperturbative power corrections. In this way, we obtain a consistent
renormalized generalization of the Bjorken sum rule for heavy--heavy%
{}~\re{Bj} and heavy--light~\re{Bj1} transitions. The former one
considerably simplifies in the small $\vartheta$ limit~\re{Bj2}.
Differentiating this sum rule, we obtain the renormalized generalizations
of Voloshin and the ``third'' sum rules~\re{Vol}.

We have analyzed nonperturbative effects in the structure functions by
considering ambiguity of perturbative corrections
to the Wilson lines to higher orders.
The non--abelian exponentiation theorem~\cite{NAexp}
together with the one--loop contribution in which the running coupling
constant is evaluated at the gluon transverse momentum~\cite{ABCMV} lead
to the improved perturbative expressions~(\ref{WC2}) and (\ref{WC21}) for
the Wilson line. These expressions contain infrared renormalon poles
at $u^*=\frac12$, $1$\dots{} on the integration contour over Borel parameter,
due to the infrared pole in $\alpha_{\rm s}(k_t)$. As a result, perturbative
expression for the Wilson line, $W_{\rm pert}$, depends on ambiguous
integration prescriptions for dealing with these singularities. Infrared
renormalons induce ambiguities into $W_{\rm pert}$ at the level of power
corrections $(\Lambda t)^{2u^*}$, at which nonperturbative
power corrections should contribute as well.

We demonstrate that the uncertainty in the structure functions induced by
the leading infrared renormalon at $u=\frac12$ in the Wilson line is
compensated by the similar uncertainty in the heavy quark pole mass.
This nicely agrees with the fact that there is no power corrections linear
in $t$. The contribution of the next infrared renormalon pole at $u=1$
to $W_{\rm pert}$ has the same dependence on the cusp angle $\vartheta$
as the nonperturbative $\CO(\Lambda^2t^2)$ correction to $W_{\rm nonpert}$
proportional to $\mu_\pi^2$. The uncertainty induced by this pole must
cancel with the ultraviolet renormalon uncertainty in the meson matrix
element $\mu_{0,\pi}^2$.
As a result, the sum of perturbative and nonperturbative
$\CO(t^2)$ contributions to the Wilson line becomes unique and it defines
the fundamental parameter $\mu_\pi^2$.

Although other nonperturbative contributions to the Wilson line are
possible, the requirement of the renormalon ambiguity cancellation
leads to the exponentiation of nonperturbative corrections~(\ref{ans})
and (\ref{ans1}).
If we cut the series in the exponents at the quadratic $\mu_\pi^2$ term,
we obtain the model nonperturbative structure functions~(\ref{toy}) and
(\ref{toy1}) describing a gaussian distribution of the heavy quark momentum
in a heavy meson. These distributions do not vanish exactly in the regions
required by the support properties. For exact fulfillment of these properties,
an infinitely many terms in the series in the exponents~(\ref{ans}),
(\ref{ans1}) are required.

{\bf Acknowledgements}.
AGG is grateful to the Royal Society and PPARC, UK, for the financial support,
and to D.~J.~Broadhurst, M.~Neubert, and N.~G.~Uraltsev for useful discussions
of various HQET problems. GPK is grateful for helpful conversations with
A.~Radyushkin and G.~Sterman. His work was supported in part by the
National Science Foundation under grant PHY9309888.

\end{document}